\UseRawInputEncoding
\documentclass{ws-ijmpa}
\usepackage[super,compress]{cite}
\usepackage{graphicx}
\begin{document}
\markboth{Z.-H. Weng}{Eight Equilibrium and continuity equations within the material media}

%
\catchline{}{}{}{}{}
%

\title{Eight equilibrium and continuity equations within the material media
}

\author{Zi-Hua Weng
}

\address{School of Aerospace Engineering, Xiamen University, Xiamen, China
\\
College of Physical Science and Technology, Xiamen University, Xiamen, China
\\
xmuwzh@xmu.edu.cn}

%

\maketitle

\begin{history}
\received{Day Month Year}
\revised{Day Month Year}
\end{history}

\begin{abstract}
  The paper aims to apply the octonions to explore the torques and forces and so forth in the electromagnetic and gravitational fields, investigating the influences of material media on the equilibrium and continuity equations. The contemporary scholars utilize the quaternions and octonions to research the electromagnetic fields and gravitational fields and so forth. In the paper, the octonions are capable of surveying the electromagnetic and gravitational fields within material media, including the octonion field strength, field source, angular momentum, torque, and force. Further, the octonion field strength and angular momentum can be combined together to become the octonion composite field strength, deducing the octonion composite torque and force and others. When the octonion composite force is equal to zero under certain circumstances, it is able to achieve eight equations independent of each other, including the force equilibrium equation, fluid continuity equation, and current continuity equation and so on. The above reveals that the material media and octonion field strength can make a contribution to the eight equilibrium and continuity equations. And it is beneficial to deepen the understanding of equilibrium and continuity equations within material media.


\end{abstract}

\ccode{PACS numbers: 03.50.De, 04.50.-h, 02.10.De, 11.10.Kk}


\section{\label{sec:level1}Introduction}

Can the material media, besides the field strength, exert an influence on some physical quantities? Is there any new equilibrium equation or continuity equation, except for three existing equilibrium or continuity equations? Can there still be some new influencing factors for the equilibrium and continuity equations? For a long time, these difficult problems have puzzled and attracted scholars. It was not until the emergence of the octonion field theory (short for the electromagnetic and gravitational theories described with the octonions) that these questions were answered in part. In the octonion field theory, when the octonion composite force (in Section 7) is equal to zero, it is able to achieve eight independent equilibrium or continuity equations simultaneously. And the material media and octonion field strength and others may make a contribution to these equilibrium and continuity equations.

In 1873, J. C. Maxwell utilized simultaneously the vector terminology and quaternions to describe the physical properties of electromagnetic fields. Subsequently, O. Heaviside and others applied the vector terminology to explore the electromagnetic fields, gravitational fields, theoretical mechanics, and hydromechanics and so forth, including the field strength, field source, angular momentum, torque, and force and so on. Further, the scholars employed the vector terminology to research the electromagnetic theory, equilibrium equations, and continuity equations and others within the electromagnetic media.

With the further development of theoretical researches described with the vector terminology, the scholars extended the research scope of electromagnetic theory from no medium to electromagnetic media. And they attempted to extend the research scope of gravitational theory from no medium to gravitational media. Furthermore, making use of the vector terminology, the scholars summarized the three famous equilibrium or continuity equations from the existing experiments, including the force equilibrium equation, fluid continuity equation, and current continuity equation.

The existing theories described with the vector terminology have made a lot of achievements, especially the electromagnetic theory and gravitational theory. However, there are also a few arduous problems in the existing electromagnetic and gravitational theories.

1) Electromagnetic medium. In terms of the existing field theories, there are two types of physical quantities, electric polarization and magnetization, for the electromagnetic media. However, there is no counterpart within the gravitational media. This situation undoubtedly obstructs the further development of the existing field theories, impeding the subsequent advancement of the gravitational theory within the gravitational medium.

2) Field source. In the existing field theories, there is merely the field source in the vacuum, but there is no field source relevant to the material media. The field theories do not take into account the contribution of the field sources related with material media on the linear momentum, energy, torque, power, and force and so forth. As a result, it is unable to explain some physical phenomena concerned with material media, confining the scope of application of the existing field theories.

3) Equilibrium equation. The existing field theories are incapable of deducing the three famous equilibrium or continuity equations, including the force equilibrium equation, fluid continuity equation, and current continuity equation. None of new equilibrium or continuity equations can be derived from the existing field theories either. Further it is unable to consider the influence of the field sources related with material media on the equilibrium and continuity equations.

The preceding analysis shows that the existing field theories have a few limitations in the exploration of several problems, relevant to the equilibrium and continuity equations and others, in the electromagnetic and gravitational media.

Presenting a striking contrast to the above is that it is able to utilize the octonions to explore a few puzzles concerned with material media, in the electromagnetic and gravitational theories, resolving some problems derived from the existing field theories.

J. C. Maxwell first applied the quaternions to study the physical properties of electromagnetic fields. Nowadays, the scholars \cite{mironov} employ the quaternions \cite{anastassiu} and octonions \cite{deleo,chanyal1} to research the electromagnetic fields \cite{demir}, gravitational fields \cite{rawat}, relativity theories \cite{morita}, curved space \cite{moffat}, quantum mechanics \cite{gogberashvili,bernevig}, dark matters \cite{furui}, astrophysical jets, strong nuclear fields \cite{furey,chanyal2}, weak nuclear fields \cite{majid,farrill}, black holes \cite{bossard}, equilibrium equations, and continuity equations \cite{tanisli,chanyal3} and others.

The algebra of octonions \cite{ward,baez} was introduced by Graves and Cayley independently. The octonions can be called as the classical octonions. What the paper discusses is the classical octonions, rather than the non-classical octonions, including hyperbolic-octonions \cite{demir1,demir2}, split-octonions \cite{demir3,demir4}, pseudo-octonions, Cartan¡¯s octonions, and others. The octonions are capable of describing simultaneously the physical properties of electromagnetic fields
and gravitational fields \cite{weng1}. Some scholars utilize the octonions \cite{tanisli1,tanisli2} to study the magnetic monopoles and massive dyons \cite{demir5,tanisli3}.

In the paper, the octonion space can be separated into a few subspaces independent of each other. One subspace is able to depict the physical properties of gravitational fields. Meanwhile, the second subspace can describe the physical properties of electromagnetic fields.

1) Gravitational medium. In the octonion field theory, it is able to deduce not only the electric polarization and magnetization for the electromagnetic media, but also the counterparts for the gravitational media. These counterparts of gravitational media are interrelated with the angular momenta. It promotes the subsequent developments of gravitational theory, exploring the angular velocities of the astrophysical jets and celestial bodies.

2) Material medium. In terms of the octonion field theory, there are not only the field sources in the vacuum, but also the field sources relevant to the material media. Consequently, it is capable of taking into account the contributions of the electromagnetic and/or gravitational media on the octonion field source, linear momentum, torque, and force and so forth, especially the energy and power.

3) Continuity equation. One can derive eight equilibrium or continuity equations, from the octonion field theory, including the force equilibrium equation, fluid continuity equation, and current continuity equation and so on. It is able to consider the influences of material media on the continuity and equilibrium equations \cite{weng2,weng3}, explaining more physical phenomena than ever before.

In the paper, the octonion field strength, field source, linear momentum, angular momentum, torque, and force can be derived from the octonion field potential and quaternion operator. Further, the octonion field strength and angular momentum can be combined together to become the octonion composite field strength. From the last, it is able to deduce the octonion composite linear momentum, angular momentum, torque, and force and so forth. All of these physical quantities take into account the contributions of material media.

\section{Octonion field source}

The octonion space $\mathbb{O}$ is able to explain the physical properties of gravitational and electromagnetic fields simultaneously. Further the octonion space can be separated into a few subspaces independent of each other, including $\mathbb{H}_g$ and $\mathbb{H}_{em}$ . The subspace $\mathbb{H}_g$ is one quaternion space, which can be applied to depict the physical properties of gravitational fields. Meanwhile, the second subspace $\mathbb{H}_{em}$ may be utilized to describe the physical properties of electromagnetic fields.

In the octonion field theory, the second subspace $\mathbb{H}_{em}$ for electromagnetic fields is independent of the quaternion space $\mathbb{H}_g$ for gravitational fields. In the quaternion space $\mathbb{H}_g$ , the radius vector is $\mathbb{R}_g = i r_0 \textbf{i}_0 + \Sigma r_k \textbf{i}_k$ , the velocity is $\mathbb{V}_g = i v_0 \textbf{i}_0 + \Sigma v_k \textbf{i}_k$, and the gravitational potential is $\mathbb{A}_g = i a_0 \textbf{i}_0 + \Sigma a_k \textbf{i}_k$ . Similarly, in the second subspace $\mathbb{H}_{em}$, the radius vector is $\mathbb{R}_e = i R_0 \textbf{I}_0 + \Sigma R_k \textbf{I}_k$ , the velocity is $\mathbb{V}_e = i V_0 \textbf{I}_0 + \Sigma V_k \textbf{I}_k$ , and the electromagnetic potential is $\mathbb{A}_e = i A_0 \textbf{I}_0 + \Sigma A_k \textbf{I}_k$ . In the octonion spaces, the octonion radius vector is $\mathbb{R} = \mathbb{R}_g + k_{eg} \mathbb{R}_e$ . The octonion velocity is $\mathbb{V} = \mathbb{V}_g + k_{eg} \mathbb{V}_e$. The octonion field potential is $\mathbb{A} = \mathbb{A}_g + k_{eg} \mathbb{A}_e$ . Herein $r_0 = v_0 t$ . $v_0$ is the speed of light, and $t$ is the time. $\textbf{r} = \Sigma r_k \textbf{i}_k$. $\textbf{i}_j$ and $\textbf{I}_j$ are two basis vectors. $\textbf{i}_0 = 1$. $\textbf{i}_k^2 = -1$. $\textbf{I}_j^2 = -1$. $\textbf{I}_j = \textbf{i}_j \circ \textbf{I}_0$. $r_j$ , $v_j$ , $a_j$ , $R_j$ , $V_j$ , and $A_j$ are all real. $\circ$ denotes the octonion multiplication. $k_{eg}$ is a coefficient, to meet the needs of dimensional homogeneity. $i$ is the imaginary unit. $j = 0, 1, 2, 3$. $k = 1, 2, 3$.

The octonion field strength $\mathbb{F}$ is defined from the octonion field potential $\mathbb{A}$ ,
\begin{eqnarray}
\mathbb{F} = \lozenge \circ \mathbb{A}  ~,
\end{eqnarray}
where $\mathbb{F} = \mathbb{F}_g + k_{eg} \mathbb{F}_e$ . $\mathbb{F}_g = \lozenge \circ \mathbb{A}_g$ , $\mathbb{F}_e = \lozenge \circ \mathbb{A}_e$ . $\mathbb{F}_g$ and $\mathbb{F}_e$ are respectively the components of octonion field strength $\mathbb{F}$ in two subspaces, $\mathbb{H}_g$ and $\mathbb{H}_{em}$ . The quaternion operator is, $\lozenge = i \textbf{i}_0 \partial_0 + \nabla$ . $\nabla = \Sigma \textbf{i}_k \partial_k$ , $\partial_j = \partial / \partial r_j$. $\mathbb{F}_g = f_0 + \textbf{f}$ , $\mathbb{F}_e = \textbf{F}_0 + \textbf{F}$ . $\textbf{f} = \Sigma f_k \textbf{i}_k$. $\textbf{F} = \Sigma F_k \textbf{I}_k$ . $\textbf{F}_0 = F_0 \textbf{I}_0$ . $f_0$ and $F_0$ are all real. $f_k$ and $F_k$ both are complex numbers.

In the above, if the gauge condition for electromagnetic fields is chosen as, $\textbf{F}_0 = - \partial_0 \textbf{A}_0 + \nabla \cdot \textbf{A} = 0$, the electromagnetic strength $\mathbb{F}_e$ will be simplified into, $\textbf{F} = i \textbf{E} / v_0 + \textbf{B}$ . The electric field intensity is, $\textbf{E} / v_0 = \partial_0 \textbf{A} + \nabla \circ \textbf{A}_0$, and the magnetic induction intensity is, $\textbf{B} = \nabla \times \textbf{A}$ . Similarly, if the gauge condition for gravitational fields is chosen as, $f_0 = - \partial_0 a_0 + \nabla \cdot \textbf{a} = 0$, the gravitational strength $\mathbb{F}_g$ will be reduced into, $\textbf{f} = i \textbf{g} / v_0 + \textbf{b}$ . The gravitational acceleration is, $\textbf{g} / v_0 = \partial_0 \textbf{a} + \nabla a_0$. And $\textbf{b} = \nabla \times \textbf{a}$ , is called as the gravitational precessional-angular-velocity temporarily. Herein $\textbf{a} = \Sigma a_k \textbf{i}_k$ . $\textbf{A} = \Sigma A_k \textbf{I}_k$. $\textbf{A}_0 = A_0 \textbf{I}_0$ .

From the octonion field strength $\mathbb{F}$ , it is able to define the octonion field source $\mu \mathbb{S}$ ,
\begin{eqnarray}
\mu \mathbb{S} = - ( i \mathbb{F} / v_0 + \lozenge )^\ast \circ \mathbb{F}  ~,
\end{eqnarray}
or
\begin{eqnarray}
- ( i \mathbb{F} / v_0 + \lozenge )^\ast \circ \mathbb{F} = \mu_g \mathbb{S}_g  + k_{eg} \mu_e \mathbb{S}_e  - ( i \mathbb{F} / v_0 )^\ast \circ \mathbb{F} ~,
\end{eqnarray}
where $\mu_g \mathbb{S}_g$ and $\mu_e \mathbb{S}_e$ are respectively the components of octonion field source $\mu \mathbb{S}$ in two subspaces, $\mathbb{H}_g$ and $\mathbb{H}_{em}$ . $\mu$ , $\mu_g$ , and $\mu_e$ are the coefficients. $\mu_g < 0$ , and $\mu_e > 0$. $\ast$ indicates the octonion conjugate. In the case of a single particle, it is obtained, $\mathbb{S}_g = m \mathbb{V}_g$, and $\mathbb{S}_e = q \mathbb{V}_e$ , by comparison with the classical field theory. $m$ is the density of mass, while $q$ is the density of electric charge. $k_{eg}^2 = \mu_g / \mu_e$ .

According to the basis vectors and coefficient $k_{eg}$ , the above can be separated into,
\begin{eqnarray}
&& \mu_g \mathbb{S}_g = - \lozenge^\ast \circ \mathbb{F}_g ~,   \label{equ:4}
\\
&& \mu_e \mathbb{S}_e = - \lozenge^\ast \circ \mathbb{F}_e ~,
\end{eqnarray}
where the former is the gravitational equations in the quaternion space, $\mathbb{H}_g$ . Meanwhile the latter is the electromagnetic equations in the second subspace, $\mathbb{H}_{em}$ .

\section{Octonion force}

From the octonion field source (Table 1), the octonion linear momentum $\mathbb{P}$ can be defined as follows,
\begin{eqnarray}
\mathbb{P} = \mu \mathbb{S} / \mu_g  ~,
\end{eqnarray}
where $\mathbb{P} = \mathbb{P}_g + k_{eg} \mathbb{P}_e$ . $\mathbb{P}_g$ and $\mathbb{P}_e$ are respectively the components of octonion linear momentum $\mathbb{P}$ in two subspaces, $\mathbb{H}_g$ and $\mathbb{H}_{em}$ . $\mathbb{P}_g = \{ \mu_g \mathbb{S}_g - ( i \mathbb{F} / v_0 )^\ast \circ \mathbb{F} \} / \mu_g $. $\mathbb{P}_e = \mu_e \mathbb{S}_e / \mu_g$. $\mathbb{P}_g = i p_0 + \textbf{p}$. $\textbf{p} = \Sigma p_k \textbf{i}_k$. $\mathbb{P}_e = i \textbf{P}_0 + \textbf{P}$. $\textbf{P} = \Sigma P_k \textbf{I}_k$. $\textbf{P}_0 = P_0 \textbf{I}_0$.

From the octonion linear momentum $\mathbb{P}$ and radius vector $\mathbb{R}$ , it is able to define the octonion angular momentum $\mathbb{L}$ as follows,
\begin{eqnarray}
\mathbb{L} = ( \mathbb{R} + k_{rx} \mathbb{X} )^\times \circ \mathbb{P}  ~ ,
\end{eqnarray}
where $\mathbb{L} = \mathbb{L}_g + k_{eg} \mathbb{L}_e$ . $\mathbb{L}_g$ and $\mathbb{L}_e$ are respectively the components of octonion angular momentum $\mathbb{L}$ in two subspaces, $\mathbb{H}_g$ and $\mathbb{H}_{em}$ . $k_{rx}$ is a coefficient, to meet the demand of dimensional homogeneity. $\mathbb{X}$ is the octonion integrating function of field potential $\mathbb{A}$ , that is, $\mathbb{A} = i \lozenge^\times \circ \mathbb{X}$ . $\mathbb{X} = \mathbb{X}_g + k_{eg} \mathbb{X}_e$ . $\mathbb{X}_g = i x_0 + \Sigma x_k \textbf{i}_k$ , $\mathbb{X}_e = i X_0 \textbf{I}_0 + \Sigma X_k \textbf{I}_k$. $\times$ denotes the complex conjugate. $\mathbb{L}_g = L_{10} + i \textbf{L}_1^i + \textbf{L}_1$ . $\mathbb{L}_e = \textbf{L}_{20} + i \textbf{L}_2^i + \textbf{L}_2$ . $\textbf{L}_1$ is the angular momentum. And $\textbf{L}_1^i$ is called as the mass moment temporarily. $\textbf{L}_2^i$ is the electric moment, while $\textbf{L}_2$ is the magnetic moment. $\textbf{L}_1 = \Sigma L_{1k} \textbf{i}_k$ . $\textbf{L}_1^i = \Sigma L_{1k}^i \textbf{i}_k$ . $\textbf{L}_{20} = L_{20} \textbf{I}_0$ . $\textbf{L}_2 = \Sigma L_{2k} \textbf{I}_k$ . $\textbf{L}_2^i = \Sigma L_{2k}^i \textbf{I}_k$ . $L_{1j}$ , $L_{1k}^i$, $L_{2j}$ , $L_{2k}^i$ , $x_j$ , and $X_j$ are all real. $k_{rx} = 1 / v_0$ .

The octonion torque $\mathbb{W}$ is defined from the octonion angular momentum,
\begin{eqnarray}
\mathbb{W} = - v_0 ( i \mathbb{F} / v_0 + \lozenge ) \circ \mathbb{L}  ~,
\end{eqnarray}
where $\mathbb{W} = \mathbb{W}_g + k_{eg} \mathbb{W}_e$ . $\mathbb{W}_g$ and $\mathbb{W}_e$ are respectively the components of octonion torque $\mathbb{W}$ in two subspaces, $\mathbb{H}_g$ and $\mathbb{H}_{em}$ . $\mathbb{W}_g = i W_{10}^i + W_{10} + i \textbf{W}_1^i + \textbf{W}_1$ . $\mathbb{W}_e = i \textbf{W}_{20}^i + \textbf{W}_{20} + i \textbf{W}_2^i + \textbf{W}_2$ . $\textbf{W}_1 = \Sigma W_{1k} \textbf{i}_k$ , $\textbf{W}_1^i = \Sigma W_{1k}^i \textbf{i}_k$. $\textbf{W}_{20} = W_{20} \textbf{I}_0$, $\textbf{W}_{20}^i = W_{20}^i \textbf{I}_0$ . $\textbf{W}_2 = \Sigma W_{2k} \textbf{I}_k$ , $\textbf{W}_2^i = \Sigma W_{2k}^i \textbf{I}_k$ . $W_{10}^i$ is the energy. $\textbf{W}_1^i$ is the torque. $\textbf{W}_{20}^i$ and $\textbf{W}_2^i$ are called as the second-energy and second-torque respectively and temporarily. $W_{1j}$ , $W_{1j}^i$ , $W_{2j}$ , and $W_{2j}^i$ are all real.

From the octonion torque, the octonion force $\mathbb{N}$ is defined as follows,
\begin{eqnarray}
\mathbb{N} = - ( i \mathbb{F} / v_0 + \lozenge ) \circ \mathbb{W}  ~,
\end{eqnarray}
where $\mathbb{N} = \mathbb{N}_g + k_{eg} \mathbb{N}_e$ . $\mathbb{N}_g$ and $\mathbb{N}_e$ are respectively the components of octonion force $\mathbb{N}$ in two subspaces, $\mathbb{H}_g$ and $\mathbb{H}_{em}$ . $\mathbb{N}_g = i N_{10}^i + N_{10} + i \textbf{N}_1^i + \textbf{N}_1$ . $\mathbb{N}_e = i \textbf{N}_{20}^i + \textbf{N}_{20} + i \textbf{N}_2^i + \textbf{N}_2$. $\textbf{N}_1 = \Sigma N_{1k} \textbf{i}_k$ , $\textbf{N}_1^i = \Sigma N_{1k}^i \textbf{i}_k$. $\textbf{N}_{20} = N_{20} \textbf{I}_0$, $\textbf{N}_{20}^i = N_{20}^i \textbf{I}_0$. $\textbf{N}_2 = \Sigma N_{2k} \textbf{I}_k$ , $\textbf{N}_2^i = \Sigma N_{2k}^i \textbf{I}_k$ . $N_{10}$ is the power. $\textbf{N}_1^i$ is the force. $\textbf{N}_{20}$ and $\textbf{N}_2^i$ are respectively called as the second-power and second-force temporarily. $N_{1j}$ , $N_{1j}^i$ , $N_{2j}$ , and $N_{2j}^i$ are all real.

\begin{table}[h]
\tbl{The physical quantities and definitions in the gravitational and electromagnetic fields without considering the contribution of material media.}
{\begin{tabular}{@{}ll@{}}
\hline\hline
physical quantity                      &   definition                                                                             \\
\hline
quaternion operator                    &   $\lozenge = i \partial_0 + \nabla$                                                     \\
octonion radius vector                 &   $\mathbb{R} = \mathbb{R}_g + k_{eg} \mathbb{R}_e$                                      \\
octonion integrating function          &   $\mathbb{X} = \mathbb{X}_g + k_{eg} \mathbb{X}_e$                                      \\
octonion field potential               &   $\mathbb{A} = i \lozenge^\times \circ \mathbb{X}$                                      \\
octonion field strength                &   $\mathbb{F} = \lozenge \circ \mathbb{A}$                                               \\
octonion field source                  &   $\mu \mathbb{S} = - ( i \mathbb{F} / v_0 + \lozenge )^\ast \circ \mathbb{F}$           \\
octonion linear momentum               &   $\mathbb{P} = \mu \mathbb{S} / \mu_g$                                                  \\
octonion angular momentum              &   $\mathbb{L} = ( \mathbb{R} + k_{rx} \mathbb{X} )^\times \circ \mathbb{P} $             \\
octonion torque                        &   $\mathbb{W} = - v_0 ( i \mathbb{F} / v_0 + \lozenge ) \circ \mathbb{L}$                \\
octonion force                         &   $\mathbb{N} = - ( i \mathbb{F} / v_0 + \lozenge ) \circ \mathbb{W}$                    \\
\hline\hline
\end{tabular}}
\end{table}

\section{Composite field source}

In the octonion space, the octonion field strength $\mathbb{F}$ and angular momentum $\mathbb{L}$ can be combined together to become the octonion composite field strength, $\mathbb{F}^+ = \mathbb{F} + k_{fl} \mathbb{L}$. $k_{fl}$ is a coefficient, to satisfy the requirement of dimensional homogeneity. The term $\mathbb{F}^+$ is the field strength within the material media.

From the octonion composite field strength $\mathbb{F}^+$ , it is able to define the octonion composite field source within the material media as follows (Table 2),
\begin{eqnarray}
\mu \mathbb{S}^+ = - ( i \mathbb{F}^+ / v_0 + \lozenge )^\ast \circ \mathbb{F}^+  ~,
\end{eqnarray}
or
\begin{eqnarray}
- ( i \mathbb{F}^+ / v_0 + \lozenge )^\ast \circ \mathbb{F}^+ = \mu_g \mathbb{S}_g^+  + k_{eg} \mu_e \mathbb{S}_e^+  - ( i \mathbb{F}^+ / v_0 )^\ast \circ \mathbb{F}^+ ~,
\end{eqnarray}
where $\mu \mathbb{S}^+ = \mu \mathbb{S} + k_{fl} \mathbb{Z}$ . The octonion physical quantity $\mathbb{Z}$ is a part of the field source, and is only related to the material media. $\mathbb{Z} = \mathbb{Z}_g + k_{eg} \mathbb{Z}_e$ . $\mathbb{Z}_g$ and $\mathbb{Z}_e$ are respectively the components of octonion physical quantity $\mathbb{Z}$ in two subspaces, $\mathbb{H}_g$ and $\mathbb{H}_{em}$ . $k_{fl} = - \mu_g$ .

From the above, it is able to achieve,
\begin{eqnarray}
&& \mu_g \mathbb{S}_g^+ = - \lozenge^\ast \circ \mathbb{F}_g^+ ~,   \label{equ:12}
\\
&& \mu_e \mathbb{S}_e^+ = - \lozenge^\ast \circ \mathbb{F}_e^+ ~,
\end{eqnarray}
where the former is the gravitational equations within gravitational media, while the latter is the electromagnetic equations within the electromagnetic media. $\mu_g \mathbb{S}_g^+ = \mu_g \mathbb{S}_g + k_{fl} \mathbb{Z}_g$ . $\mu_e \mathbb{S}_e^+ = \mu_e \mathbb{S}_e + k_{fl} \mathbb{Z}_e$ . $\mathbb{F}^+ = \mathbb{F}_g^+ + k_{eg} \mathbb{F}_e^+$ . $\mathbb{F}_g^+$ and $\mathbb{F}_e^+$ are respectively the components of octonion composite field strength $\mathbb{F}^+$ in two subspaces, $\mathbb{H}_g$ and $\mathbb{H}_{em}$ . The two gauge conditions are chosen as, $\textbf{F}_0^+ = 0$, and $f_0^+ = 0$.

\section{Composite angular momentum}

From the octonion composite field source, $\mathbb{S}^+$ , it is capable of defining the octonion composite linear momentum within the material media,
\begin{eqnarray}
\mathbb{P}^+ = \mu \mathbb{S}^+ / \mu_g  ~,
\end{eqnarray}
where $\mathbb{P}^+ = \mathbb{P}_g^+ + k_{eg} \mathbb{P}_e^+$ . $\mathbb{P}_g^+ = \{ \mu_g \mathbb{S}_g^+ - ( i \mathbb{F}^+ / v_0 )^\ast \circ \mathbb{F}^+ \} / \mu_g $ . $\mathbb{P}_e^+ = \mu_e \mathbb{S}_e^+ / \mu_g$ . $\mathbb{P}_g^+$ and $\mathbb{P}_e^+$ are respectively the components of octonion composite linear momentum $\mathbb{P}^+$ in two subspaces, $\mathbb{H}_g$ and $\mathbb{H}_{em}$ . $\mathbb{P}_g^+ = i p_0^+ + \textbf{p}^+$ . $\textbf{p}^+ = \Sigma p_k^+ \textbf{i}_k$. $\mathbb{P}_e^+ = i \textbf{P}_0^+ + \textbf{P}^+$. $\textbf{P}^+ = \Sigma P_k^+ \textbf{I}_k$ . $\textbf{P}_0^+ = P_0^+ \textbf{I}_0$ .

From the octonion composite linear momentum, $\mathbb{P}^+$ , and composite radius vector, $\mathbb{R}^+ = \mathbb{R} + k_{rx} \mathbb{X}$ , one can define the octonion composite angular momentum within the material media,
\begin{eqnarray}
\mathbb{L}^+ = ( \mathbb{R}^+ )^\times \circ \mathbb{P}^+  ~ ,
\end{eqnarray}
where $\mathbb{L}^+ = \mathbb{L}_g^+ + k_{eg} \mathbb{L}_e^+$ . $\mathbb{L}_g^+ = L_{10}^+ + i \textbf{L}_1^{i+} + \textbf{L}_1^+$ . $\mathbb{L}_e^+ = \textbf{L}_{20}^+ + i \textbf{L}_2^{i+} + \textbf{L}_2^+$ . $\mathbb{L}_g^+$ and $\mathbb{L}_e^+$ are respectively the components of octonion composite angular momentum $\mathbb{L}^+$ in two subspaces, $\mathbb{H}_g$ and $\mathbb{H}_{em}$ . $\mathbb{R}^+ = \mathbb{R}_g^+ + k_{eg} \mathbb{R}_e^+$. $\mathbb{R}_g^+ = \mathbb{R}_g + k_{rx} \mathbb{X}_g$. $\mathbb{R}_e^+ = \mathbb{R}_e + k_{rx} \mathbb{X}_e$. $r_j^+ = r_j + k_{rx} x_j$ , $R_j^+ = R_j + k_{rx} X_j$ . $\textbf{r}^+ = \Sigma r_k^+ \textbf{i}_k$ . $\textbf{R}_0^+ = R_0^+ \textbf{I}_0$ , $\textbf{R}^+ = \Sigma R_k^+ \textbf{I}_k$.

In the above, $\textbf{L}_1^+$ is the angular momentum within the material media. $\textbf{L}_1^{i+}$ is the mass moment within the material media. $\textbf{L}_2^{i+}$ is the electric moment within the material media. $\textbf{L}_2^+$ is the magnetic moment within the material media.

Some terms in the above can be written as follows,
\begin{eqnarray}
&& L_{10}^+ = r_0^+ p_0^+ + \textbf{r}^+ \cdot \textbf{p}^+ + k_{eg}^2 ( \textbf{R}_0^+ \circ \textbf{P}_0^+ + \textbf{R}^+ \cdot \textbf{P}^+ ) ~,   \\
&& \textbf{L}_1^{i+} = p_0^+ \textbf{r}^+ - r_0^+ \textbf{p}^+  +  k_{eg}^2 ( \textbf{R}^+ \circ \textbf{P}_0^+ - \textbf{R}_0^+ \circ \textbf{P}^+ )  ~,  \\
&& \textbf{L}_1^+ = \textbf{r}^+ \times \textbf{p}^+ + k_{eg}^2 \textbf{R}^+ \times \textbf{P}^+ ~, \\
&& \textbf{L}_{20}^+ = r_0^+ \textbf{P}_0^+ + \textbf{r}^+ \cdot \textbf{P}^+ + p_0^+ \textbf{R}_0^+ + \textbf{R}^+ \cdot \textbf{p}^+ ~,  \\
&& \textbf{L}_2^{i+} = - r_0^+ \textbf{P}^+ + \textbf{r}^+ \circ \textbf{P}_0^+  - \textbf{R}_0^+ \circ \textbf{p}^+ + p_0^+ \textbf{R}^+  ~,  \\
&& \textbf{L}_2^+ = \textbf{r}^+ \times \textbf{P}^+ + \textbf{R}^+ \times \textbf{p}^+ ~,
\end{eqnarray}
where $\textbf{L}_1^+ = \Sigma L_{1k}^+ \textbf{i}_k$ . $\textbf{L}_1^{i+} = \Sigma L_{1k}^{i+} \textbf{i}_k$ . $\textbf{L}_{20}^+ = L_{20}^+ \textbf{I}_0$ . $\textbf{L}_2^+ = \Sigma L_{2k}^+ \textbf{I}_k$ . $\textbf{L}_2^{i+} = \Sigma L_{2k}^{i+} \textbf{I}_k$. $L_{1j}^+$ , $L_{1k}^{i+}$ , $L_{2j}^+$ , and $L_{2k}^{i+}$ are all real.

\section{Composite torque}

From the octonion composite angular momentum $\mathbb{L}^+$ , it is able to define the octonion composite torque within the material media as follows (Table 3),
\begin{eqnarray}
\mathbb{W}^+ = - v_0 ( i \mathbb{F}^+ / v_0 + \lozenge ) \circ \mathbb{L}^+  ~,
\end{eqnarray}
where $\mathbb{W}^+ = \mathbb{W}_g^+ + k_{eg} \mathbb{W}_e^+$ . $\mathbb{W}_g^+ = i W_{10}^{i+} + W_{10}^+ + i \textbf{W}_1^{i+} + \textbf{W}_1^+$ . $\mathbb{W}_e^+ = i \textbf{W}_{20}^{i+} + \textbf{W}_{20}^+ + i \textbf{W}_2^{i+} + \textbf{W}_2^+$ . $\mathbb{W}_g^+$ and $\mathbb{W}_e^+$ are respectively the components of octonion composite torque $\mathbb{W}^+$ in two subspaces, $\mathbb{H}_g$ and $\mathbb{H}_{em}$ . $\textbf{W}_1^+ = \Sigma W_{1k}^+ \textbf{i}_k$, $\textbf{W}_1^{i+} = \Sigma W_{1k}^{i+} \textbf{i}_k$. $\textbf{W}_{20}^+ = W_{20}^+ \textbf{I}_0$ , $\textbf{W}_{20}^{i+} = W_{20}^{i+} \textbf{I}_0$ . $\textbf{W}_2^+ = \Sigma W_{2k}^+ \textbf{I}_k$ , $\textbf{W}_2^{i+} = \Sigma W_{2k}^{i+} \textbf{I}_k$ . $W_{1j}^+$ , $W_{1j}^{i+}$ , $W_{2j}^+$ , and $W_{2j}^{i+}$ are all real.

In the above, $W_{10}^{i+}$ is the energy within the material media. And it consists of not only the proper energy, kinetic energy, potential energy, and work, but also the interacting energy between the electric field with electric moment, and the interacting energy between the magnetic field with magnetic moment and others within the material media. $\textbf{W}_1^{i+}$ is the torque within the material media, including the torque produced by the applied force and others. $\textbf{W}_{20}^{i+}$ and $\textbf{W}_2^{i+}$ are respectively the second-energy and second-torque within the material media.

Some terms in the above can be written as follows,
\begin{eqnarray}
W_{10}^{i+}  = &&  ( \textbf{g}^+ \cdot \textbf{L}_1^{i+} / v_0 - \textbf{b}^+ \cdot \textbf{L}_1^+ )
                     - v_0 (  \partial_0 L_{10}^+ +  \nabla \cdot \textbf{L}_1^{i+} )
                     \nonumber \\
                     &&
                     + k_{eg}^2 ( \textbf{E}^+ \cdot \textbf{L}_2^{i+} / v_0 - \textbf{B}^+ \cdot \textbf{L}_2^+ ) ~,
\\
W_{10}^+  = && ( \textbf{b}^+ \cdot \textbf{L}_1^{i+} + \textbf{g}^+ \cdot \textbf{L}_1^+ / v_0 ) - v_0 ( \nabla \cdot \textbf{L}_1^+ )
                     \nonumber \\
                     && + k_{eg}^2 ( \textbf{B}^+ \cdot \textbf{L}_2^{i+} + \textbf{E}^+ \cdot \textbf{L}_2^+ / v_0 ) ~,
\\
\textbf{W}_1^{i+}  = && ( \textbf{g}^+ \times \textbf{L}_1^{i+} / v_0 - L_{10}^+ \textbf{b}^+ - \textbf{b}^+ \times \textbf{L}_1^+ )
                     - v_0 (  \partial_0 \textbf{L}_1^+ +  \nabla \times \textbf{L}_1^{i+} )
                     \nonumber \\
                     && + k_{eg}^2 ( \textbf{E}^+ \times \textbf{L}_2^{i+} / v_0 - \textbf{B}^+ \circ \textbf{L}_{20}^+ - \textbf{B}^+ \times \textbf{L}_2^+ ) ~,
\\
\textbf{W}_1^+  = && ( \textbf{g}^+ L_{10}^+ / v_0 + \textbf{g}^+ \times \textbf{L}_1^+ / v_0 + \textbf{b}^+ \times \textbf{L}_1^{i+} )
                     \nonumber \\
                     && - v_0 ( - \partial_0 \textbf{L}_1^{i+} + \nabla L_{10}^+ +  \nabla \times \textbf{L}_1^+ )
                     \nonumber \\
                     && + k_{eg}^2 ( \textbf{E}^+ \circ \textbf{L}_{20}^+ / v_0 + \textbf{E}^+ \times \textbf{L}_2^+ / v_0 + \textbf{B}^+ \times \textbf{L}_2^{i+} ) ~,
\\
%
%
\textbf{W}_{20}^{i+} = && ( \textbf{g}^+ \cdot \textbf{L}_2^{i+} / v_0 - \textbf{b}^+ \cdot \textbf{L}_2^+ )
                     + ( \textbf{E}^+ \cdot \textbf{L}_1^{i+} / v_0 - \textbf{B}^+ \cdot \textbf{L}_1^+ )
                     \nonumber \\
                     && - v_0 ( \partial_0 \textbf{L}_{20}^+ + \nabla \cdot \textbf{L}_2^{i+} ) ~,
\\
\textbf{W}_{20}^+ = && ( \textbf{b}^+ \cdot \textbf{L}_2^{i+} + \textbf{g}^+ \cdot \textbf{L}_2^+ / v_0 ) - v_0 ( \nabla \cdot \textbf{L}_2^+ )
                     \nonumber \\
                     && + ( \textbf{B}^+ \cdot \textbf{L}_1^{i+} + \textbf{E}^+ \cdot \textbf{L}_1^+ / v_0 ) ~,
\\
\textbf{W}_2^{i+} = && ( \textbf{g}^+ \times \textbf{L}_2^{i+} / v_0 - \textbf{b}^+ \circ \textbf{L}_{20}^+ - \textbf{b}^+ \times \textbf{L}_2^+ )
                     \nonumber \\
                     && + ( \textbf{E}^+ \times \textbf{L}_1^{i+} / v_0 - L_{10}^+ \textbf{B}^+ - \textbf{B}^+ \times \textbf{L}_1^+ )
                     \nonumber \\
                     && - v_0 ( \partial_0 \textbf{L}_2^+ + \nabla \times \textbf{L}_2^{i+} ) ~,
\\
\textbf{W}_2^+ = && ( \textbf{g}^+ \circ \textbf{L}_{20}^+ / v_0 + \textbf{g}^+ \times \textbf{L}_2^+ / v_0 + \textbf{b}^+ \times \textbf{L}_2^{i+} )
                     \nonumber \\
                     && + ( L_{10}^+ \textbf{E}^+ / v_0 + \textbf{E}^+ \times \textbf{L}_1^+ / v_0 + \textbf{B}^+ \times \textbf{L}_1^{i+} )
                     \nonumber \\
                     && - v_0 ( - \partial_0 \textbf{L}_2^{i+} + \nabla \circ \textbf{L}_{20}^+ + \nabla \times \textbf{L}_2^+ ) ~.
\end{eqnarray}

Further, the energy $W_{10}^{i+}$ within the material media can be written approximately to,
\begin{eqnarray}
W_{10}^{i+} \approx  && - \{ v_0 p_0^+ + v_0 p_0^+ ( \nabla \cdot \textbf{r} ) \}
                        - \{ v_0 ( \partial_0 \textbf{r}) \cdot \textbf{p}^+ + v_0 \textbf{r} \cdot \partial_0 \textbf{p}^+ \}
                        \nonumber \\
                        && - \{ ( p_0^+ a_0 + \textbf{a} \cdot \textbf{p}^+ ) + k_{eg}^2 ( \textbf{A}_0 \circ \textbf{P}_0^+ + \textbf{A} \cdot \textbf{P}^+ ) \}
                        \nonumber \\
                        && +  k_{eg}^2 \{ (\textbf{E}^+ / v_0 ) \cdot ( \textbf{r} \circ \textbf{P}_0^+ ) - \textbf{B}^+ \cdot ( \textbf{r} \times \textbf{P}^+ ) \}
                        \nonumber \\
                        && + \{ ( \textbf{g}^+ / v_0 ) \cdot ( p_0^+ \textbf{r} ) - \textbf{b}^+ \cdot \textbf{L}_1^+ \} ~,
\end{eqnarray}
where $\textbf{r}^+ \approx \textbf{r}$. The term, $- \{ v_0 p_0^+ + v_0 p_0^+ ( \nabla \cdot \textbf{r} ) \} = k_p v_0 p_0^+$ , covers the proper energy $v_0 s_0^+$ and others within the material media. The term, $- \{ v_0 ( \partial_0 \textbf{r}) \cdot \textbf{p}^+ + v_0 \textbf{r} \cdot \partial_0 \textbf{p}^+ \}$ , is the sum of the kinetic energy and the work produced by the applied force within the material media. The term, $- \{ ( p_0^+ a_0 + \textbf{a} \cdot \textbf{p}^+ ) + k_{eg}^2 ( \textbf{A}_0 \circ \textbf{P}_0^+ + \textbf{A} \cdot \textbf{P}^+ ) \}$ , is the potential energy of gravitational and electromagnetic fields within the material media. The term, $\{ k_{eg}^2  (\textbf{E}^+ / v_0 ) \cdot ( \textbf{r} \circ \textbf{P}_0^+ ) \}$, is the interacting energy between the electric field intensity with the electric moment within the material media. The term, $\{ - k_{eg}^2 \textbf{B}^+ \cdot ( \textbf{r} \times \textbf{P}^+ ) \}$ , is the interacting energy between the magnetic flux density with the magnetic moment within the material media. The term, $( - \textbf{b}^+ \cdot \textbf{L}_1^+ )$, is the interacting energy between the gravitational precessional-angular-velocity with the angular momentum within the material media. For the case of $k = 3$, the term $( W_{10}^{i+} / 2 )$ is the conventional energy within the material media. $k_p = (k - 1)$ is the coefficient, with $k$ being the dimension of radius vector $\textbf{r}$.

Similarly, the torque $-\textbf{W}_1^{i+}$ within the material media can be written approximately to,
\begin{eqnarray}
\textbf{W}_1^{i+} \approx &&  p_0^+ \textbf{g}^+ \times \textbf{r} / v_0 - v_0 \textbf{r} \times \partial_0 \textbf{p}^+
                              \nonumber \\
                              &&
                              + \{ - ( \textbf{r} \cdot \textbf{p}^+ ) \textbf{b}^+ - \textbf{b}^+ \times ( \textbf{r} \times \textbf{p}^+ )
                              + \textbf{a} \times \textbf{p}^+ \}
                              \nonumber \\
                              && + k_{eg}^2 \{ \textbf{E}^+ \times ( \textbf{r} \circ \textbf{P}_0^+ ) / v_0
                              -  \textbf{B}^+ \times ( \textbf{r} \times \textbf{P}^+ ) \} ~,
\end{eqnarray}
where $\textbf{r}^+ \approx \textbf{r}$. The term $( - p_0^+ \textbf{g}^+ \times \textbf{r} / v_0 )$ is the torque produced by the gravity within the material media. The term $( v_0 \textbf{r} \times \partial_0 \textbf{p}^+ )$ is the torque produced by the inertial force within the material media. The term, $\{ - k_{eg}^2 \textbf{E}^+ \times ( \textbf{r} \circ \textbf{P}_0 )^+ / v_0 \}$ , is the torque caused by the electric fields and electric moments within the material media. The term, $\{ k_{eg}^2 \textbf{B}^+ \times ( \textbf{r} \times \textbf{P}^+ ) \}$, is the torque caused by the magnetic fields and magnetic moments within the material media.

It's worth noting that the octonion composite field strength, composite field source, composite linear momentum, composite angular momentum, composite torque, and composite force are related with the material media in the octonion spaces. These six physical quantities within the material media are respectively different from the counterparts without the material media in Table 1. However, none of octonion radius vector, field potential, and integrating function relates to the material media. The three physical quantities remain unchanged, which are respectively the same as those in Table 1.

\section{Composite force}

From the octonion composite torque $\mathbb{W}^+$ , one can define the octonion composite force within the material media,
\begin{eqnarray}
\mathbb{N}^+ = - ( i \mathbb{F}^+ / v_0 + \lozenge ) \circ \mathbb{W}^+  ~,
\end{eqnarray}
where $\mathbb{N}^+ = \mathbb{N}_g^+ + k_{eg} \mathbb{N}_e^+$ . $\mathbb{N}_g^+ = i N_{10}^{i+} + N_{10}^+ + i \textbf{N}_1^{i+} + \textbf{N}_1^+$ . $\mathbb{N}_e^+ = i \textbf{N}_{20}^{i+} + \textbf{N}_{20}^+ + i \textbf{N}_2^{i+} + \textbf{N}_2^+$ . $\mathbb{N}_g^+$ and $\mathbb{N}_e^+$ are respectively the components of octonion composite force $\mathbb{N}^+$ in two subspaces, $\mathbb{H}_g$ and $\mathbb{H}_{em}$ . $\textbf{N}_1^+ = \Sigma N_{1k}^+ \textbf{i}_k$, $\textbf{N}_1^{i+} = \Sigma N_{1k}^{i+} \textbf{i}_k$. $\textbf{N}_{20}^+ = N_{20}^+ \textbf{I}_0$ , $\textbf{N}_{20}^{i+} = N_{20}^{i+} \textbf{I}_0$. $\textbf{N}_2^+ = \Sigma N_{2k}^+ \textbf{I}_k$ , $\textbf{N}_2^{i+} = \Sigma N_{2k}^{i+} \textbf{I}_k$ . $N_{1j}^+$ , $N_{1j}^{i+}$ , $N_{2j}^+$ , and $N_{2j}^{i+}$ are all real.

In the above, $N_{10}^+$ is the power within the material media, which relates to the fluid continuity equation. $\textbf{N}_1^{i+}$ is the force within the material media, which deals with the force equilibrium equation. $\textbf{N}_{20}^+$ is the second-power within the material media, which relates with the current continuity equation. $\textbf{N}_2^{i+}$ is the second-force within the material media, which associates with the second-force equilibrium equation.

Some terms in the above can be written as follows,
\begin{eqnarray}
N_{10}^{i+}  = && ( \textbf{g}^+ \cdot \textbf{W}_1^{i+} / v_0 - \textbf{b}^+ \cdot \textbf{W}_1^+ ) / v_0
                       +  k_{eg}^2 ( \textbf{E}^+ \cdot \textbf{W}_2^{i+} / v_0 - \textbf{B}^+ \cdot \textbf{W}_2^+ ) / v_0
                       \nonumber \\
                       && - ( \partial_0 W_{10}^+ + \nabla \cdot \textbf{W}_1^{i+} ) ~ ,
\\
N_{10}^+  = && ( \textbf{g}^+ \cdot \textbf{W}_1^+ / v_0 + \textbf{b}^+ \cdot \textbf{W}_1^{i+} ) / v_0
                       + k_{eg}^2 ( \textbf{E}^+ \cdot \textbf{W}_2^+ / v_0 + \textbf{B}^+ \cdot \textbf{W}_2^{i+} ) / v_0
                       \nonumber \\
                       && + ( \partial_0 W_{10}^{i+} -  \nabla \cdot \textbf{W}_1^+ ) ~  ,
\\
\textbf{N}_1^{i+} = && k_{eg}^2 ( \textbf{E}^+ \circ \textbf{W}_{20}^{i+} / v_0 + \textbf{E}^+ \times \textbf{W}_2^{i+} / v_0
                       - \textbf{B}^+ \circ \textbf{W}_{20}^+ - \textbf{B}^+ \times \textbf{W}_2^+ ) / v_0
                       \nonumber \\
                       && + ( W_{10}^{i+} \textbf{g}^+ / v_0 + \textbf{g}^+ \times \textbf{W}_1^{i+} / v_0
                       - W_{10}^+ \textbf{b}^+ - \textbf{b}^+ \times \textbf{W}_1^+ ) / v_0
                       \nonumber \\
                       && - ( \partial_0 \textbf{W}_1^+ + \nabla W_{10}^{i+} + \nabla \times \textbf{W}_1^{i+} ) ~ ,
\\
\textbf{N}_1^+ = && k_{eg}^2 ( \textbf{E}^+ \circ \textbf{W}_{20}^+ / v_0 + \textbf{E}^+ \times \textbf{W}_2^+ / v_0
                       + \textbf{B}^+ \circ \textbf{W}_{20}^{i+} + \textbf{B}^+ \times \textbf{W}_2^{i+} ) / v_0
                       \nonumber \\
                       && + ( W_{10}^+ \textbf{g}^+ / v_0 + \textbf{g}^+ \times \textbf{W}_1^+ / v_0
                       + W_{10}^{i+} \textbf{b}^+ + \textbf{b}^+ \times \textbf{W}_1^{i+} ) / v_0
                       \nonumber \\
                       && + ( \partial_0 \textbf{W}_1^{i+} - \nabla W_{10}^+ - \nabla \times \textbf{W}_1^+ )  ~ ,
\\
\textbf{N}_{20}^{i+} = && ( \textbf{g}^+ \cdot \textbf{W}_2^{i+} / v_0 - \textbf{b}^+ \cdot \textbf{W}_2^+ ) / v_0
                       - ( \partial_0 \textbf{W}_{20}^+ + \nabla \cdot \textbf{W}_2^{i+} )
                       \nonumber \\
                       && + ( \textbf{E}^+ \cdot \textbf{W}_1^{i+} / v_0 - \textbf{B}^+ \cdot \textbf{W}_1^+ ) / v_0    ~,
\\
\textbf{N}_{20}^+ = && ( \textbf{g}^+ \cdot \textbf{W}_2^+ / v_0 + \textbf{b}^+ \cdot \textbf{W}_2^{i+} ) / v_0
                       + ( \partial_0 \textbf{W}_{20}^{i+} - \nabla \cdot \textbf{W}_2^+ )
                       \nonumber \\
                       && + ( \textbf{E}^+ \cdot \textbf{W}_1^+ / v_0 + \textbf{B}^+ \cdot \textbf{W}_1^{i+} ) / v_0    ~,
\\
\textbf{N}_2^{i+} = && ( \textbf{g}^+ \circ \textbf{W}_{20}^{i+} / v_0 + \textbf{g}^+ \times \textbf{W}_2^{i+} / v_0
                       - \textbf{b}^+ \circ \textbf{W}_{20}^+ - \textbf{b}^+ \times \textbf{W}_2^+ ) / v_0
                       \nonumber \\
                       && + ( W_{10}^{i+} \textbf{E}^+ / v_0 + \textbf{E}^+ \times \textbf{W}_1^{i+} / v_0
                       - W_{10}^+ \textbf{B}^+ - \textbf{B}^+ \times \textbf{W}_1^+ ) / v_0
                       \nonumber \\
                       && - ( \partial_0 \textbf{W}_2^+ + \nabla \circ \textbf{W}_{20}^{i+} + \nabla \times \textbf{W}_2^{i+} )  ~,
\\
\textbf{N}_2^+ = && ( \textbf{g}^+ \circ \textbf{W}_{20}^+ / v_0 + \textbf{g}^+ \times \textbf{W}_2^+ / v_0
                       + \textbf{b}^+ \circ \textbf{W}_{20}^{i+} + \textbf{b}^+ \times \textbf{W}_2^{i+} ) / v_0
                       \nonumber \\
                       && + ( W_{10}^+ \textbf{E}^+ / v_0 + \textbf{E}^+ \times \textbf{W}_1^+ / v_0
                       + W_{10}^{i+} \textbf{B}^+ + \textbf{B}^+ \times \textbf{W}_1^{i+} ) / v_0
                       \nonumber \\
                       && + ( \partial_0 \textbf{W}_2^{i+} - \nabla \circ \textbf{W}_{20}^+ - \nabla \times \textbf{W}_2^+  )  ~.
\end{eqnarray}

When the octonion composite force is equal to zero, one can achieve simultaneously eight continuity or equilibrium equations independent of each other, including the fluid continuity equation, force equilibrium equation, precession equilibrium equation, torque continuity equation, current continuity equation, second-force equilibrium equation, second-precession equilibrium equation, and second-torque continuity equation within the material media (Table 4).

\begin{table}[h]
\tbl{The octonion composite physical quantities in the gravitational and electromagnetic fields considering the contribution of material media.}
{\begin{tabular}{@{}ll@{}}
\hline\hline
physical quantity                               &   definition                                                                                 \\
\hline
composite field strength                        &   $\mathbb{F}^+ = \mathbb{F} + k_{fl} \mathbb{L}$                                            \\
composite field source                          &   $\mu \mathbb{S}^+ = - ( i \mathbb{F}^+ / v_0 + \lozenge )^\ast \circ \mathbb{F}^+$         \\
composite linear momentum                       &   $\mathbb{P}^+ = \mu \mathbb{S}^+ / \mu_g$                                                  \\
composite angular momentum                      &   $\mathbb{L}^+ = ( \mathbb{R} + k_{rx} \mathbb{X} )^\times \circ \mathbb{P}^+ $             \\
composite octonion torque                       &   $\mathbb{W}^+ = - v_0 ( i \mathbb{F}^+ / v_0 + \lozenge ) \circ \mathbb{L}^+$              \\
composite octonion force                        &   $\mathbb{N}^+ = - ( i \mathbb{F}^+ / v_0 + \lozenge ) \circ \mathbb{W}^+$                  \\
\hline\hline
\end{tabular}}
\end{table}

\subsection{Fluid continuity equation}

If the octonion composite field strength is relatively weak, the power $N_{10}^+$ within the material media can be written approximately as,
\begin{eqnarray}
N_{10}^+ / k_p \approx  && \partial_0 ( p_0^+ v_0 ) - \nabla \cdot ( \textbf{p}^+ v_0 )
+ L_{10}^+ ( \textbf{b}^+ \cdot \textbf{b}^+ - \textbf{g}^+ \cdot \textbf{g}^+ / v_0^2 ) / ( v_0 k_p )
\nonumber \\
&&
+ k_{eg}^2 L_{10}^+ ( \textbf{B}^+ \cdot \textbf{B}^+ - \textbf{E}^+ \cdot \textbf{E}^+ / v_0^2 ) / ( v_0 k_p )
 \nonumber \\
                              &&
+ k_{eg}^2 \textbf{E}^+ \cdot \textbf{P}^+ / v_0 + \textbf{g}^+ \cdot \textbf{p}^+ / v_0 ~,
\label{equ:42}
\end{eqnarray}
where $W_{10}^{i+} \approx k_p p_0^+ v_0$ , $\textbf{W}_1^+ \approx k_p \textbf{p}^+ v_0$ , $\textbf{W}_2^+ \approx k_p \textbf{P}^+ v_0$ .

When $N_{10}^+ = 0$ , the fluid continuity equation within the material media can be derived from the above. Further, in the extreme condition there is no octonion composite field strength, it will be simplified into the mass continuity equation within the material media,
\begin{equation}
\partial_0 p_0^+ - \nabla \cdot \textbf{p}^+ = 0 ~.
\end{equation}

\subsection{Force equilibrium equation}

If the octonion composite field strength is comparative weak, the force $\textbf{N}_1^{i+}$ within the material media can be written approximately as,
\begin{eqnarray}
\textbf{N}_1^{i+} / k_p \approx  && - \partial_0 (\textbf{p}^+ v_0)  + p_0^+ \textbf{g}^+ / v_0
- \textbf{b}^+ \times \textbf{p}^+ - \nabla (p_0^+ v_0)
 \nonumber \\
                              &&
+ L_{10}^+ ( \textbf{g}^+ \times \textbf{b}^+ + k_{eg}^2 \textbf{E}^+ \times \textbf{B}^+ ) / ( v_0^2 k_p )
 \nonumber \\
                              &&
+ k_{eg}^2 ( \textbf{E}^+ \circ \textbf{P}_0^+ / v_0 - \textbf{B}^+ \times \textbf{P}^+ ) ~,
\label{equ:44}
\end{eqnarray}
where $\textbf{W}_{20}^{i+} \approx k_p \textbf{P}_0^+ v_0$ . $\textbf{W}_2^{i+} \approx \textbf{v} \times \textbf{P}^+ $ . $( p_0^+ \textbf{g}^+ / v_0 )$ is the force within the material media. $ \partial_0 ( - \textbf{p}^+ v_0)$ is the inertial force within the material media. $ \{k_{eg}^2 ( \textbf{E}^+ \circ \textbf{P}_0^+ / v_0 - \textbf{B}^+ \times \textbf{P}^+ ) \}$ is the electromagnetic force within the material media. $\nabla (p_0^+ v_0)$ is the energy gradient within the material media. $\{ L_{10}^+ (  k_{eg}^2 \textbf{E}^+ \times \textbf{B}^+ ) / ( v_0^2 k_p ) \}$ is in direct proportion to the electromagnetic momentum within the material media.

When $\textbf{N}_1^{i+} = 0$, the force equilibrium equation within the material media can be derived from the above, including some existing and new force terms of electromagnetic and gravitational fields within the material media. Further, if some tiny terms are neglected, it will be reduced into the force equilibrium equation within the material media,
\begin{eqnarray}
- \partial_0 (\textbf{p}^+ v_0)  + p_0^+ \textbf{g}^+ / v_0 - \nabla (p_0^+ v_0)
+ k_{eg}^2 ( \textbf{E}^+ \circ \textbf{P}_0^+ / v_0 - \textbf{B}^+ \times \textbf{P}^+ ) = 0 ~.
\end{eqnarray}

\subsection{Current continuity equation}

If the octonion composite field strength is comparative weak, the second-power $\textbf{N}_{20}^+$ within the material media can be written approximately as,
\begin{eqnarray}
\textbf{N}_{20}^+ / k_p  \approx && \partial_0 ( \textbf{P}_0^+ v_0 ) - \nabla \cdot ( \textbf{P}^+ v_0 )
 \nonumber \\
                              &&
+ \textbf{g}^+ \cdot \textbf{P}^+ / v_0 + \textbf{E}^+ \cdot \textbf{p}^+ / v_0
 \nonumber \\
                              &&
+ ( \textbf{b}^+ \cdot \textbf{b}^+ + \textbf{B}^+ \cdot \textbf{B}^+ ) \textbf{L}_{20}^+ / ( v_0 k_p ) ~,
\label{equ:46}
\end{eqnarray}
where the electromagnetic strength and gravitational strength within the material media both make a contribution to the above.

When $\textbf{N}_{20}^+ = 0$, it is able to derive the current continuity equation from the above within the material media, which includes the interacting term, $( \textbf{g}^+ \cdot \textbf{P}^+ + \textbf{E}^+ \cdot \textbf{p}^+ )$ , between the gravitational field and electromagnetic field within the material media. It implies that the current continuity equation may be disturbed by some influencing factors relevant to the material media.

Further, under the extreme condition there is no octonion composite field strength, the above will be degenerated into the current continuity equation within the material media,
\begin{equation}
\partial_0 \textbf{P}_0^+ - \nabla \cdot \textbf{P}^+ = 0 ~.
\end{equation}

\subsection{Second-force equilibrium equation}

In case $\textbf{N}_2^{i+} = 0$ , and the octonion composite field strength is comparative weak, we can achieve the second-force equilibrium equation within the material media as follows,
\begin{eqnarray}
0 = ( \textbf{g}^+ \circ \textbf{W}_{20}^{i+} / v_0 - \textbf{b}^+ \circ \textbf{W}_{20}^+ + W_{10}^{i+} \textbf{E}^+ / v_0 - W_{10}^+ \textbf{B}^+ ) / v_0 - \partial_0 \textbf{W}_2^+ ~ .
\end{eqnarray}

The above means that the current derivatives and charge gradients must satisfy the requirement of the second-force equilibrium equation. The electromagnetic strength and gravitational strength and others will exert a significant impact on the derivative of magnetic moment or the current derivative within the material media.

If the derivatives, $\partial_t \textbf{B}^+$ , $\partial_t \textbf{E}^+$ , $\partial_t \textbf{g}^+$ , and $\partial_t \textbf{b}^+$ and so forth can be neglected, the second-force equilibrium equation within the material media can be approximately simplified as,
\begin{eqnarray}
0 = && ( \partial_t \textbf{S}^+ + v_0 \nabla \circ \textbf{S}_0^+ ) - ( \textbf{g}^+ \circ \textbf{S}_0^+ / v_0 - \textbf{b}^+ \times \textbf{S}^+ )
 \nonumber \\
                              &&
- ( \mu_g / \mu_e ) ( p_0^+ \textbf{E}^+ / v_0 - \textbf{B}^+ \times \textbf{p}^+ ) ~,
\end{eqnarray}
where $\partial_t \textbf{S}^+$ is the derivative of electric current within the material media, while $\nabla \circ \textbf{S}_0^+ / v_0$ is the gradient of electric charge within the material media. $\textbf{S}^+$ is the density of electric current within the material media, and $q^+$ is the density of electric charge within the material media. $\textbf{S}_0^+ = S_0^+ \textbf{I}_0$ . $S_0^+ = q^+ v_0$ . $\partial_t = \partial / \partial t$ .

a) When the electromagnetic strength $( \textbf{E}^+ , \textbf{B}^+ )$ within the material media is comparatively weak, while the gravitational strength $( \textbf{g}^+ , \textbf{b}^+ )$ within the material media is comparatively strong, and the coefficient $( \mu_g / \mu_e )$ is quite small, the second-force equilibrium equation within the material media can be degenerated to,
\begin{eqnarray}
0 = ( \partial_t \textbf{S}^+ + v_0 \nabla \circ \textbf{S}_0^+ ) - ( \textbf{g}^+ \circ \textbf{S}_0^+ / v_0 - \textbf{b}^+ \times \textbf{S}^+ )  ~.
\end{eqnarray}

The above states that the comparatively strong gravitational strength will make a contribution to the second-force equilibrium equation within the material media.

b) When the gravitational strength $( \textbf{g}^+ , \textbf{b}^+ )$ within the material media is comparatively weak, while the electromagnetic strength $( \textbf{E}^+ , \textbf{B}^+ )$ within the material media is comparatively strong, the second-force equilibrium equation within the material media can be reduced to,
\begin{eqnarray}
0 = ( \partial_t \textbf{S}^+ + v_0 \nabla \circ \textbf{S}_0^+ ) - ( \mu_g / \mu_e ) ( p_0^+ \textbf{E}^+ / v_0 - \textbf{B}^+ \times \textbf{p}^+ ) ~.
\end{eqnarray}

The above means that the comparatively strong electromagnetic strength may exert an influence on the second-force equilibrium equation within the material media

c) When the gravitational strength and electromagnetic strength within the material media both are comparatively weak, the second-force equilibrium equation within the material media can be simplified into,
\begin{eqnarray}
0 =  \partial_t \textbf{S}^+ + v_0 \nabla \circ \textbf{S}_0^+  ~.
\end{eqnarray}

The above implies that the current derivative and charge gradient within the material media are closely correlated, in the case of comparatively weak field strength within the material media. And the second-force equilibrium equation within the material media should be satisfied between two of them.

According to the above, the charge gradients are able to induce the current derivatives within the material media. This inference is validated by the experiments relevant to the transport of droplets \cite{deng,corum}. Meanwhile the current derivatives are capable of inducing the charge gradients within the material media \cite{ober,cyli}.

\subsection{Precession equilibrium equation}

In case $\textbf{N}_1^+ = 0$, it is able to achieve the precession equilibrium equation within the material media, deducing the angular velocities of precession for charged or neutral particles within the material media. And the field strength and angular momentum/electromagentic moment and others may exert an influence on the precession equilibrium equation within the material media.

a) Torque-derivative term. In case two terms, $\partial_0 \textbf{W}_1^{i+}$ and $\nabla \times \textbf{W}_1^+$ , play a major role in the precessional motion within the material media, meanwhile other tiny terms can be neglected, the precession equilibrium equation, $\textbf{N}_1^+ = 0$, within the material media will be degenerated into,
\begin{eqnarray}
\partial_0 \textbf{W}_1^{i+} - k_p m^+ k v_0 \overrightarrow{\omega}_p =  0     ~ ,
\label{equ:18}
\end{eqnarray}
where $\overrightarrow{\omega}_p$ is the angular velocity of precession within the material media. $k_p = (k - 1)$ is one coefficient, and $k$ is the spatial dimension of vector $\textbf{r}$ , or that of linear velocity $\textbf{v}_p$ .

The above states that the angular velocity of precession is related with the physical quatities within the material media, including the spatial dimension, mass, and torque-derivative term and so forth. And the torque-derivative term, $\partial_0 \textbf{W}_1^{i+}$ , within the material media will produce the angular velocity of precession, even if there is no field strength.

b) Magnetic field. If two terms, $k_{eg}^2 \textbf{B}^+ \circ \textbf{W}_{20}^{i+} / v_0$ and $\nabla \times \textbf{W}_1^+$ , play a major role in the precessional motion within the material media, meanwhile other tiny terms can be neglected, the precession equilibrium equation, $\textbf{N}_1^+ = 0$, within the material media will be simplified into,
\begin{eqnarray}
k_{eg}^2 \textbf{B}^+ \circ \textbf{W}_{20}^{i+} / v_0 - \nabla \times \textbf{W}_1^+  =  0  ~ ,
\end{eqnarray}
further the above will be reduced into,
\begin{eqnarray}
q^+ \textbf{B}^+ \circ \textbf{\emph{I}}_0 - k m^+ \overrightarrow{\omega}_p = 0    ~ ,
\label{equ:20}
\end{eqnarray}
where the angular velocity of precession within the material media is, $\overrightarrow{\omega}_{p(1)} = q^+ \textbf{B}^+ \circ \textbf{\emph{I}}_0 / m^+$, when $k = 1$. The angular velocity of precession within the material media is, $\overrightarrow{\omega}_{p(2)} = q^+ \textbf{B}^+ \circ \textbf{\emph{I}}_0 / (2 m^+)$ , when $k = 2$. And the angular velocity of precession within the material media is, $\overrightarrow{\omega}_{p(3)} = q^+ \textbf{B}^+ \circ \textbf{\emph{I}}_0 / (3 m^+)$ , when $k = 3$.

The above means that the magnetic flux density $\textbf{B}^+$ will induce the angular velocity of precession within the material media, revolving around the direction, $\textbf{B}^+ \circ \textbf{\emph{I}}_0$ , for the charged objects. The inference can be applied to explicate the angular velocity of Larmor precession (see Ref.[30]) within the material media.

c) Electric field. When two terms, $k_{eg}^2 ( \textbf{E}^+ \circ \textbf{W}_{20}^+ ) / v_0^2$ and $\nabla \times \textbf{W}_1^+$ , play a major role in the precessional motion within the material media, and other tiny terms can be neglected, the precession equilibrium equation, $\textbf{N}_1^+ = 0$, within the material media can be degraded into,
\begin{eqnarray}
k_{eg}^2  \textbf{E}^+ \circ \textbf{W}_{20}^+ / v_0^2  - \nabla \times \textbf{W}_1^+  = 0  ~ .
\label{equ:21}
\end{eqnarray}

The above reveals that the electric field intensity $\textbf{E}^+$ will generate the angular velocity of precession, with the orientation of precession, $\textbf{E}^+ \circ \textbf{\emph{I}}_0$ , for the charged objects within the material media. And it can be utilized to unpuzzle some precessional phenomena of charged particles relevant to the Stark effect within the material media.

d) Gravitational precessional-angular-velocity. When two terms, $W_{10}^{i+} \textbf{b}^+ / v_0$ and $\nabla \times \textbf{W}_1^+$ , play a major role in the precessional motion within the material media, and other tiny terms can be neglected, the precession equilibrium equation, $\textbf{N}_1^+ = 0$, within the material media can be degenerated into,
\begin{eqnarray}
W_{10}^{i+} \textbf{b}^+ / v_0 - \nabla \times \textbf{W}_1^+ = 0   ~ ,
\label{equ:22}
\end{eqnarray}
further the above can be simplified into,
\begin{eqnarray}
\textbf{b}^+ - k \overrightarrow{\omega}_p = 0    ~ ,
\label{equ:23}
\end{eqnarray}
where the angular velocity of precession within the material media is, $\overrightarrow{\omega}_{p(1)} = \textbf{b}^+ $, when $k = 1$. The angular velocity of precession within the material media is, $\overrightarrow{\omega}_{p(2)} = \textbf{b}^+ / 2$, when $k = 2$. And the angular velocity of precession within the material media is, $\overrightarrow{\omega}_{p(3)} = \textbf{b}^+ / 3$, when $k = 3$.

The above states that the gravitational precessional-angular-velocity $\textbf{b}^+$ will induce the angular velocity of precession, with the orientation of precession, $\textbf{b}^+$ , for the neutral objects within the material media. And it can be applied to account for the dynamic properties of the astrophysical jets within the material media, including the precession, rotation, and collimation and so forth.

e) Gravitational acceleration. In case two terms, $ W_{10}^+ \textbf{g}^+ / v_0^2$ and $\nabla \times \textbf{W}_1^+$ , play a major role in the precessional motion within the material media, and other tiny terms can be neglected, the precession equilibrium equation, $\textbf{N}_1^+ = 0$ , within the material media can be degraded into,
\begin{eqnarray}
W_{10}^+ \textbf{g}^+ / v_0^2 - \nabla \times \textbf{W}_1^+ = 0    ~ .
\label{equ:24}
\end{eqnarray}

The above states that the gravitational acceleration will result in the angular velocity of precession, with the orientation of precession, $\textbf{g}^+$ , for the neutral objects within the material media. And it can be applied to explain some precessional phenomena of neutral particles in the gravitational field $\textbf{g}^+$ within the material media.

\begin{table}[h]
\tbl{Comparison of some composite physical quantities between the gravitational field and electromagnetic field within the material media.}
{\begin{tabular}{@{}lll@{}}
\hline\hline
physical quantity                  &   gravitational field                                            &   electromagnetic field                                  \\
\hline
composite field strength           &   gravitational acceleration, $\textbf{g}^+$                     &   electric field intensity, $\textbf{E}^+$               \\
                                   &   gravitational precessional-angular-velocity, $\textbf{b}^+$    &   magnetic induction intensity, $\textbf{B}^+$           \\
composite field source             &   mass, $s_0^+$                                                  &   electric charge, $\textbf{S}_0^+$                      \\
                                   &   linear momentum, $\textbf{s}^+$                                &   electric current, $\textbf{S}^+$                       \\
composite angular momentum         &   dot product, $L_{10}^+$                                        &   dot product, $\textbf{L}_{20}^+$                       \\
                                   &   angular momentum, $\textbf{L}_1^+$                             &   magnetic moment, $\textbf{L}_2^+$                      \\
                                   &   mass moment, $\textbf{L}_1^{i+}$                               &   electric moment, $\textbf{L}_2^{i+}$                   \\
composite torque                   &   divergence of angular momentum, $W_{10}^+$                     &   divergence of magnetic moment, $\textbf{W}_{20}^+$     \\
                                   &   energy, $W_{10}^{i+}$                                          &   second-energy, $\textbf{W}_{20}^{i+}$                  \\
                                   &   curl of angular momentum, $\textbf{W}_1^+$                     &   curl of magnetic moment, $\textbf{W}_2^+$              \\
                                   &   torque, $\textbf{W}_1^{i+}$                                    &   second-torque, $\textbf{W}_2^{i+}$                     \\
composite force                    &   power, $N_{10}^+$                                              &   second-power, $\textbf{N}_{20}^+$                      \\
                                   &   torque divergence, $N_{10}^{i+}$                               &   second-torque divergence, $\textbf{N}_{20}^{i+}$       \\
                                   &   torque derivative, $\textbf{N}_1^+$                            &   second-torque derivative, $\textbf{N}_2^+$             \\
                                   &   force, $\textbf{N}_1^{i+}$                                     &   second-force, $\textbf{N}_2^{i+}$                      \\
\hline\hline
\end{tabular}}
\end{table}

\section{Experiment proposal}

When the octonion composite force $\mathbb{N}^+$ is equal to zero, it is capable of achieving eight equations independent of each other, including the fluid continuity equation, force equilibrium equation, current continuity equation, precession equilibrium equation, and second-force equilibrium equation and others. These eight independent equilibrium/continuity equations within the material media are derived from one single octonion equation, $\mathbb{N}^+ = 0$. It shows that the eight independent equations are essentially the same and equally important within the material media.

According to the definition of octonion composite force $\mathbb{N}^+$ , it is found that the octonion composite field strength $\mathbb{F}^+$ will make a contribution to the octonion composite force $\mathbb{N}^+$ within the material media. As a result, the octonion composite field strength $\mathbb{F}^+$ must have an influence on the equilibrium/continuity equations within the material media. In other words, the octonion composite field strength $\mathbb{F}^+$ is able to exert an impact on the fluid continuity equation, force equilibrium equation, and current continuity equation and others within the material media.

1) Fluid continuity equation. On the basis of the existing experiments to validate the fluid continuity equation, one can improve the existing experimental schemes to some extent, verifying the contribution of material media and octonion field strength to the fluid continuity equation. According to Eq.(\ref{equ:42}), the Joule heat, $(- k_{eg} \textbf{E}^+ \cdot \textbf{P}^+ / v_0 )$, in the ultra-strong electric fields will make a significant contribution to the fluid continuity equation of the ordinary fluids or magneto-fluids.

2) Force equilibrium equation. Based on the existing experiments to verify the force equilibrium equation, we may amend the existing experimental schemes to a certain extent, reckoning the contribution of the material media and octonion field strength to the force equilibrium equation. According to Eq.(\ref{equ:44}), the energy gradient, $\nabla ( p_0^+ v_0 )$, in the ultra-strong magnetic fields will exert a significant impact on the force equilibrium equation of some particles.

3) Current continuity equation. On the basis of the existing experiments to confirm the current continuity equation, it is able to meliorate the existing experimental schemes to a certain extent, surveying the contribution of the material media and octonion field strength to the current continuity equation. According to Eq.(\ref{equ:46}), the interacting term, $( \textbf{E}^+ \cdot \textbf{p}^+ / v_0 )$ , in the ultra-strong electric fields will have an influence on the current continuity equation of the ordinary fluids or magneto-fluids.

4) Second-force equilibrium equation. Based upon the existing experiments (see Ref.[32]) to research the second-force equilibrium equation, one may improve the existing experimental schemes to a certain extent, reckoning the contribution of the material media and octonion field strength to the second-force equilibrium equation.

According to the second-force equilibrium equation, the current derivatives are able to induce the charge gradients within the material media, achieveing the bidirectional transports of droplets. And the octonion field strength and angular momentum/electromagentic moment will have an influence on the second-force equilibrium equation relevant to the superconducting currents and charge gradients within the material media.

5) Precession equilibrium equation. On the basis of the existing experiments (for instance, Electron Spin Resonance and others) to study the sample materials with unpaired electrons, one can utilize the Crystal Lattice Vibration methods (such as, X-ray, $\gamma$-ray, or neutron inelastic scattering) to vibrate the crystal lattices.

As a result, it is able to generate the three-dimensional precessional motions of unpaired electrons surrounding the crystal lattices within the sample materials (see Ref.[30]), achieving the absorption spectrum of electromagnetic waves caused by the three-dimensional precessional motions within the material media.

The examination of these suggested experiments will help to deepen the understanding of the physical properties of equilibrium and continuity equations.

\begin{table}[h]
\tbl{From the octonion composite force equation, $\mathbb{N}^+ = 0$, it is able to achieve simultaneously eight continuity or equilibrium equations independent of each other within the material media, including four continuity equations and four equilibrium equations.}
{\begin{tabular}{@{}lll@{}}
\hline\hline
definition                                &  equilibrium/continuity equation                    &   subspace           \\
\hline
$N_{10}^+ = 0$                            &  fluid continuity equation                          &   $\mathbb{H}_g$     \\
$N_{10}^{i+} = 0$                         &  torque continuity equation                         &   $\mathbb{H}_g$     \\
$\textbf{N}_1^+ = 0$                      &  precession equilibrium equation                    &   $\mathbb{H}_g$     \\
$\textbf{N}_1^{i+} = 0$                   &  force equilibrium equation                         &   $\mathbb{H}_g$     \\
$\textbf{N}_{20}^+ = 0$                   &  current continuity equation                        &   $\mathbb{H}_{em}$  \\
$\textbf{N}_{20}^{i+} = 0$                &  second-torque continuity equation                  &   $\mathbb{H}_{em}$  \\
$\textbf{N}_2^+ = 0$                      &  second-precession equilibrium equation             &   $\mathbb{H}_{em}$  \\
$\textbf{N}_2^{i+} = 0$                   &  second-force equilibrium equation                  &   $\mathbb{H}_{em}$  \\
\hline\hline
\end{tabular}}
\end{table}

\section{Conclusions and discussions}

In the paper, the quaternion space $\mathbb{H}_g$ for gravitational fields is independent of the second subspace $\mathbb{H}_{em}$ for electromagnetic fields. The two independent subspaces can be considered to be perpendicular to each other. In other words, the octonion spaces can be applied to describe simultaneously the physical properties of electromagnetic and gravitational fields.

The application of the algebra of octonions is able to study the physical quantities of electromagnetic and gravitational fields, including the octonion field potential, field strength, field source, linear momentum, angular momentum, torque, and force. However, all of these physical quantities do not consider the contribution coming from the material media. For instance, the gravitational equations Eq.(\ref{equ:4}), derived from the definition of octonion field source, do not take account of the contribution of gravitational media. Therefore, it is necessary to introduce the new physical quantity from an appropriate point of view, considering the contribution of material media.

In the octonion spaces, the octonion field strength $\mathbb{F}$ and angular momentum $\mathbb{L}$ can be combined together to become one new physical quantity, that is the octonion composite field strength $\mathbb{F}^+$. And it includes the electromagnetic strength and gravitational strength with the material media. Further, from the octonion composite field strength, it is capable of deducing the octonion composite field source, composite linear momentum, composite angular momentum, composite torque, and composite force. Apparently, these composite physical quantities take into account the influence of electromagnetic and gravitational media.

With the expansion of the definition of octonion composite field source, one can achieve the electromagnetic equations within the electromagnetic media, and the gravitational equations Eq.(\ref{equ:12}) within the gravitational media. From the point of view of the algebra of octonions, the existing electromagnetic equations are a mixture of two parts. The first part is the electromagnetic equations without the electromagnetic media, while the second part is with the electromagnetic media.

When the octonion composite force $\mathbb{N}^+$ is equal to zero, it is capable of attaining simultaneously eight equations independent of each other, including the fluid continuity equation, force equilibrium equation, precession equilibrium equation, torque continuity equation, current continuity equation, second-force equilibrium equation, second-precession equilibrium equation, and second-torque continuity equation within the material media. The analysis states that the material media and octonion field strength can directly make a contribution to the eight equilibrium or continuity equations.

It is noteworthy that the paper only discusses a few simple cases about the octonion composite physical quantities relevant to the electromagnetic and gravitational fields within material media, but the study is clearly revealed the influence of material media and octonion field strength on the physical quantities. The correlative inferences can be applied to research several physical properties of electromagnetic and gravitational media, including the investigations relevant to the angular velocities of astrophysical jets and heavenly bodies. In the future studies, it is going to further explore the influence factors of continuity and equilibrium equations theoretically, especially the research relevant to the bidirectional transports of droplets. Meanwhile, we plan to verify the impact effect of these influence factors experimentally, deepening the understanding of the physical properties of continuity and equilibrium equations.

\section*{Acknowledgments}
The author is indebted to the anonymous referees for their valuable comments on the previous manuscripts. This project was supported partially by the National Natural Science Foundation of China under grant number 60677039.


\end{document}